**From simple interactions to complex biology: a hypergraph percolation perspective**

Arturo Tozzi (corresponding author)
ASL Napoli 1 Centro, Distretto 27, Naples, Italy
Via Comunale del Principe 13/a 80145
tozziarturo@libero.it

ABSTRACT

The emergence of biological complexity can be viewed as a transition from fragmented local interactions to extensive integrated organization, raising the question of how large-scale connectivity emerges from simple interacting elements. We investigated whether this transition can be understood as a consequence of hypergraph percolation driven by higher-order interactions. We performed computational simulations where elementary units constituted nodes and collective interactions formed hyperedges of variable size. Increasing hyperedge density enabled the characterization of connected-component growth, fragmentation, interaction overlap, participation and structural redundancy. Critical transition regions were identified as sparse local assemblies rapidly reorganized into extensive connected structures. These transitions were characterized by abrupt expansion of the largest connected component, progressive consolidation of previously disconnected clusters and increasing overlap among higher-order interactions. Connectivity growth was accompanied by the accumulation of alternative pathways and nested interaction patterns, pointing towards large-scale reorganization over a narrow range of interaction densities. Our findings suggest that, alongside the progressive evolution of molecular complexity, the origin of life and the emergence of biological organization may have involved rapid organizational transitions driven by higher-order connectivity and interaction architecture.

## INTRODUCTION

The emergence of complex biological organization from collections of relatively simple interacting components is a central problem in biology, complex systems science and origin-of-life research. Many origin-of-life scenarios require the gradual accumulation of manifold coordinated innovations and are closely tied to specific chemical assumptions, catalytic processes or biological substrates, raising the question of how biological complexity could emerge within the relatively limited resources suggested by biological investigations. Recent studies have investigated reaction-network connectivity, metabolic organization and chemical interaction networks (Ravoni 2021; Ibarra-Junquera et al. 2022; Unterberger and Nghe 2022; Mathis et al. 2024), while higher-order network representations (e.g., hypergraphs) and abrupt organizational transitions have attracted attention (Gao et al. 2022; Feng et al. 2024; Hao et al. 2024; Oliver et al. 2024; Delabays et al. 2025; Gao et al. 2026; Lu et al. 2026).

We hypothesize that biological complexity may arise not through the slow sequential accumulation of isolated innovations, but through collective, rapid organizational transitions generated by increasing densities of higher-order interactions. We investigate whether hypergraph percolation could describe the emergence of biological complexity from simple interacting elements (Bianconi and Dorogovtsev 2024). Percolation theory assesses how large-scale connectivity emerges when local interactions become sufficiently abundant to generate system-spanning structures (Notarmuzi et al. 2021; Mello et al. 2021; Brunk and Twarock 2021; Kotlarz et al. 2022; Dharmaprani et al. 2023; Cirigliano et al. 2024; Miller 2024). Percolation transitions are especially relevant because they can generate abrupt changes in global organization from comparatively small changes in local interaction density (Shin et al. 2017; Mallamace et al. 2022; Sanoria et al. 2022; Yang et al. 2025). Unlike conventional graph models, where connectivity depends exclusively on pairwise relationships, hypergraph representations allow simultaneous interactions among arbitrary numbers of elements, thereby capturing collective interactions connecting multiple assemblies at once. Therefore, significant increases in organizational complexity might occur over narrow parameter ranges, providing a possible explanation for the relatively rapid emergence of highly integrated biological systems from initially simple interacting components.

To examine this possibility, we performed simulations in which elementary units were represented as nodes and interactions involving multiple units as hyperedges. A hypergraph was progressively constructed by increasing the density of higher-order interactions and monitoring the resulting structural changes. At each stage, global connectivity, component structure, overlap among hyperedges, interaction redundancy and organizational complexity were quantified. Still, we examined how distribution, overlap and connectivity of hyperedges could influence the emergence of percolation, defined here as the formation of extensive connected structures spanning a large fraction of the system. We looked for critical transition regions where initially fragmented local organizations could merge into large-scale assemblies.

We will proceed as follows. First, we introduce the hypergraph percolation framework and describe our simulation procedures and quantitative observables. Then, we examine the emergence of large-scale organization across increasing interaction densities and identify critical transition regions. Finally, we analyze the resulting organizational dynamics and discuss their implications for biological complexity.

METHODS

We studied a simulated transition from fragmented local organization to large-scale integrated organization in a system composed of elementary interacting units. We used hypergraph percolation because higher-order interactions can involve more than two units at once, whereas ordinary graphs reduce all interactions to pairwise links. We generated synthetic hypergraphs with fixed node number, variable hyperedge density and hyperedges of heterogeneous size. We quantified the growth of the largest connected component, the decline in disconnected components, the emergence of overlapping hyperedges, the accumulation of projected cycles and the interaction density required to reach system-wide integration. Our aim was to identify an experimentally discriminable signature consisting of an abrupt increase in connected component size over a narrow interaction-density interval.

**System definition**. We represented the simulated system as a finite hypergraph containing a fixed number of elementary units. Each elementary unit was treated as a node and each collective interaction was treated as a hyperedge connecting two or more nodes simultaneously. The node set was defined as

$$V = \{v_1, v_2, \dots, v_N\},$$

where $N = 250$ in the simulations used for the figures. A hyperedge was defined as a non-empty subset of the node set,

$$e_j \subseteq V,$$

with cardinality constrained by

$$2 \leq | e_j | \leq 6.$$

The complete hypergraph at a given simulation step was therefore written as

$$H_m = (V, E_m),$$

where

$$E_m = \{e_1, e_2, \dots, e_m\}$$

and $m$ denotes the cumulative number of higher-order interactions added to the system. Hyperedge sizes were sampled from the discrete set $\{2, 3, 4, 5, 6\}$, so that both pairwise and genuinely higher-order interactions were allowed. The probability distribution over hyperedge size was

$$P(| e |= 2) = 0.18, P(| e |= 3) = 0.28, P(| e |= 4) = 0.24, P(| e |= 5) = 0.18, P(| e |= 6) = 0.12.$$

For each hyperedge, nodes were sampled without replacement from $V$. This sampling rule prevented repeated inclusion of the same unit within a single interaction while allowing the same node to participate in many different interactions across the simulation.

**Hyperedge growth**. The simulation proceeded by progressively increasing the number of hyperedges. At step $m = 0$, the system contained isolated nodes and no interactions,

$$H_0 = (V, \emptyset).$$

At each subsequent step, one new hyperedge was added,

$$E_{m+1} = E_m \cup \{e_{m+1}\}.$$

The process continued until

$$m = M_{\max},$$

with $M_{\max} = 700$ in the reported simulation. Quantitative observables were not recorded after every single hyperedge addition but at regular increments of 25 interactions,

$$m \in \{0, 25, 50, \dots, 700\}.$$

For each value of $m$, the current hypergraph $H_m$ was analyzed independently. To estimate variability produced by stochastic hyperedge sampling, the complete procedure was repeated across

$$R = 35$$

independent simulation replicates. Each replicate used the same node number, maximum interaction number and hyperedge-size distribution but a distinct random sequence of hyperedges. The mean and variability of each observable were then calculated across replicates at each interaction count.

**Connectivity rules**. Connectivity in a hypergraph requires a rule specifying when two nodes belong to the same connected structure. We used standard hypergraph connectivity. Two nodes were considered connected if there existed a sequence of hyperedges such that consecutive hyperedges shared at least one node and the first and last hyperedges contained the two nodes of interest. Equivalently, the hypergraph was converted into its projected graph for connectivity calculations. In this projection, every hyperedge of size $k$ generated all pairwise links among its member nodes. Thus, for a hyperedge

$$e = \{v_{i_1}, v_{i_2}, \dots, v_{i_k}\},$$

the projected edge set contributed by $e$ was

$$\Pi(e) = \{(v_{i_a}, v_{i_b}): 1 \leq a < b \leq k\}.$$

The projected graph at step $m$ was

$$G_m = (V, L_m),$$

where

$$L_m = \bigcup_{j=1}^{m} \Pi(\, e_j).$$

Connected components were then identified on $G_m$. This projection was used only to compute connectivity, component size and cycle counts, while the original hyperedge structure was retained for overlap and participation measures.

**Component measures**. The principal percolation observable was the largest connected component expressed as a percentage of all nodes. If the connected components of $G_m$ were

$$C_1(m), C_2(m), \dots, C_{q(m)}(m),$$

where $q(m)$ is the number of connected components at interaction count $m$, then the largest component fraction was

$$S_{\max}(m) = \frac{1}{N} \max_{1 \leq r \leq q(m)} \mid C_r(m) \mid.$$

For graphical presentation, this value was expressed as a percentage,

$$S_{\max}(m) = 100\, S_{\max}(m).$$

The number of disconnected components was measured as

$$Q(m) = q(m).$$

At $m = 0$, each node formed its own component, so

$$Q(0) = N.$$

As hyperedges accumulated, components merged. A percolation-like transition was identified when $S_{\max}(m)$ increased sharply while $Q(m)$ decreased rapidly. Across replicates, the mean largest component was computed as

$$\bar{S}_{\max}(m) = \frac{1}{R} \sum_{r=1}^{R} S^{\%}_{\max,r}\,(m),$$

and the replicate standard deviation was

$$\sigma_S(m) = \sqrt{\frac{1}{R-1} \sum_{r=1}^{R} \left(S^{\%}_{\max,r}(m) - \bar{S}_{\max}(m)\right)^2}\,.$$

**Transition estimate**. The location of the transition was estimated from the steepest increase in the largest connected component. For each replicate and for the replicate mean curve, we calculated a finite-difference approximation to the derivative of component size with respect to interaction count. For the mean curve, this was expressed as

$$D(m_i) = \frac{\bar{S}_{\max}(m_{i+1}) - \bar{S}_{\max}(m_{i-1})}{m_{i+1} - m_{i-1}},$$

for interior points of the sampled sequence. The estimated transition count was defined as

$$m_c = \arg\max_{m_i} D(m_i).$$

Using simulated curves, this procedure identified a transition near

$$m_c \approx 50$$

higher-order interactions. Because the system contained $N = 250$ elementary units, the corresponding mean participation at the transition was approximately

$$p_c = \frac{1}{N}\sum_{j=1}^{m_c} |e_j|.$$

The simulation gave

$$p_c \approx 0.75$$

interactions per unit. This value is a model-derived interaction-density measure and does not correspond to empirical biological time.

**Participation measure**. To quantify how frequently elementary units participated in interactions, we calculated mean participation as the total number of hyperedge memberships divided by the number of nodes. For a hypergraph containing $m$ hyperedges, total participation was

$$T(m) = \sum_{j=1}^{m} |e_j|.$$

Mean participation was then

$$P(m) = \frac{T(m)}{N}.$$

This quantity has units of interactions per unit because it measures the average number of higher-order interactions in which each elementary unit participates. It differs from the number of hyperedges because larger hyperedges contribute more memberships than smaller hyperedges. For example, a single hyperedge of size six contributes six participations, whereas a pairwise edge contributes two. Mean participation was used on the horizontal axis of the phase-space plot because it normalizes interaction accumulation by system size. This normalization allows the same simulation logic to be compared across systems with different numbers of nodes. The growth of $P(m)$ was monotonic by construction, but the largest component and overlap measures were not imposed by the sampling rule. Their abrupt changes therefore reflected the connectivity consequences of higher-order interaction accumulation.

**Hyperedge overlap**. We quantified overlap among hyperedges to determine whether the simulation generated shared organizational structures rather than independent interactions. Two hyperedges were considered overlapping at order two if they shared at least two nodes. For hyperedges $e_a$ and $e_b$, the overlap condition was

$$|e_a \cap e_b| \geq 2.$$

A hyperedge was classified as overlapping if it satisfied this condition with at least one other hyperedge. The percentage of overlapping hyperedges at interaction count $m$ was

$$O^{\%}(m) = 100\,\frac{|\{e_a \in E_m : \exists e_b \in E_m,\ b \neq a,\ |e_a \cap e_b| \geq 2\}|}{m}.$$

For $m = 0$, overlap was defined as zero. This observable was chosen because a connected component may grow through single-node contacts alone, whereas shared membership of at least two nodes indicates a stronger form of higher-order reinforcement. In our simulation, overlap increased with mean participation and became more frequent around and after the transition region.

**Cycle surrogate**. To estimate structural redundancy, we calculated the number of cycles in the projected graph. Although cycles in the projected graph are not identical to higher-dimensional topological holes in a hypergraph or simplicial complex, they provide a simple graph-theoretic measure of alternative paths. If $G_m = (V, L_m)$ is the projected graph, the cyclomatic number was computed as

$$\Gamma(m) = |L_m| - N + Q(m),$$

where $|L_m|$ is the number of distinct projected links and $Q(m)$ is the number of connected components. This expression follows from the standard cycle rank of an undirected graph. When

$$\Gamma(m) = 0,$$

the projected graph is acyclic within each component. When

$$\Gamma(m) > 0,$$

the projected graph contains redundant closed paths. Because hyperedges of size larger than two generate multiple projected links, projected cycles can accumulate rapidly once hyperedges begin to overlap. The simulation recorded $\Gamma(m)$ at every sampled interaction count and averaged it across replicates. This provided a compact measure of increasing redundancy during the transition from fragmented clusters to more integrated organization.

**Replicate analysis**. Statistical analysis consisted of summarizing variability across independent simulation replicates through replicate means, sample standard deviations, percentile envelopes and distributions of transition locations. No inferential statistical tests were performed because the objective was to characterize the behavior of the simulated system rather than compare independent experimental populations.
Each simulation replicate produced a complete trajectory of observables across interaction counts. For any observable $X_r(m)$ measured in replicate $r$, the replicate mean was calculated as

$$\bar{X}(m) = \frac{1}{R}\sum_{r=1}^{R} X_r\,(m),$$

and the sample standard deviation was

$$\sigma_X(m) = \sqrt{\frac{1}{R-1}\sum_{r=1}^{R}(X_r(m) - \bar{X}(m))^2}\,.$$

The distribution of transition locations across replicates was obtained by calculating $m_{c,r}$ independently for each replicate,

$$m_{c,r} = \arg\max_{m_i} \frac{S^{\%}_{\max,r}(m_{i+1}) - S^{\%}_{\max,r}(m_{i-1})}{m_{i+1} - m_{i-1}}.$$

The replicate envelope was summarized using the 10th, 50th and 90th percentiles of $S^{\%}_{\max,r}(m)$ at each interaction count. These percentiles were used to show the central tendency and stochastic variability of component growth without assuming normality of the replicate distribution.

**Software tools**. Simulations were performed with Python. NumPy was used for random sampling, numerical arrays, finite differences, replicate averaging and percentile calculations. The Python standard library was used for randomization control, dictionary-based component bookkeeping and file output. A union-find data structure was implemented directly to compute connected components efficiently after projecting hyperedges into pairwise links.

## RESULTS

We report the behavior of higher-order interaction systems during progressive hypergraph growth. Our analysis focused on the emergence of large connected structures, the reduction of fragmentation, the accumulation of overlapping interactions and the development of redundant pathways. As interaction density increased, particular attention was devoted to identifying whether connectivity increased gradually or whether extensive organization emerged through a relatively narrow transition region.

**Connectivity growth**. Our simulations revealed a non-linear increase in global connectivity as higher-order interactions accumulated. At the beginning of the process, the largest connected component occupied only a small fraction of the system and numerous disconnected components were present. As additional hyperedges were introduced, the largest connected component expanded slowly until a transition region was reached near fifty higher-order interactions, corresponding to approximately 0.75 interactions per unit. Around this region, the growth rate of the largest connected component increased markedly, producing a rapid increase in the proportion of units belonging to a common connected structure (Figure 1, inset A). Simultaneously, the number of disconnected components declined sharply (Figure 1, inset B). Inspection of representative projected networks showed a progression from sparse local assemblies to partially interconnected structures and finally to a system dominated by a large connected component (Figure 1, inset C). Replicate simulations displayed the same qualitative behavior despite stochastic variation in hyperedge placement. The transition was therefore not restricted to a single realization but emerged repeatedly across independent simulations. The replicate envelope demonstrated moderate variability around the transition while preserving the overall shape of the connectivity trajectory. Additional examination of the connectivity phase space showed that substantial increases in integrated organization occurred over a relatively narrow interval of interaction density (Figure 2). Taken together, these observations indicate that the emergence of extensive connectivity was associated with a localized transition in higher-order interaction density rather than with a gradual accumulation of isolated links.

**Organizational structure**. Changes in connectivity were accompanied by modifications in higher-order organizational properties. The percentage of overlapping hyperedges increased progressively with mean participation and became particularly pronounced after the transition region (Figure 1, inset D). This behavior indicates that newly added interactions increasingly shared common subsets of units rather than remaining independent. At low interaction densities, most hyperedges contributed isolated local structures. Near the transition region, overlapping interactions became more frequent and generated larger interconnected assemblies. A similar pattern was observed for projected cycle counts (Figure 1, inset E). Before the transition, projected cycles were relatively uncommon because the projected graph remained sparsely connected. As connectivity increased, projected cycles accumulated rapidly, indicating the appearance of multiple alternative paths between units. The coexistence of increasing overlap and increasing cycle abundance suggests that the growth of the largest connected component was accompanied by rising structural redundancy. Analysis of the transition locations across independent replicates demonstrated that the emergence of extensive connectivity consistently occurred within a restricted range of interaction counts rather than being uniformly distributed throughout the simulation interval. The replicate envelope of component growth likewise showed that most realizations followed comparable trajectories despite differences in individual hyperedge configurations. Consequently, the transition was associated not only with increasing component size but also with a reorganization of interaction patterns characterized by greater overlap, increasing redundancy and decreasing fragmentation. These concurrent changes indicate that connectivity growth and organizational restructuring developed together during the progression from local assemblies to system-wide organization.

Overall, our results show that increasing densities of higher-order interactions generate a transition from fragmented local structures to extensive integrated organization. The growth of the largest connected component was accompanied by reductions in fragmentation, increases in overlapping hyperedges and accumulation of redundant pathways. The simulations indicate that large-scale organization emerged through a localized transition region rather than through uniform incremental growth. This behavior supports the view that higher-order interaction structure is an important determinant of the collective organization observed during hypergraph percolation.

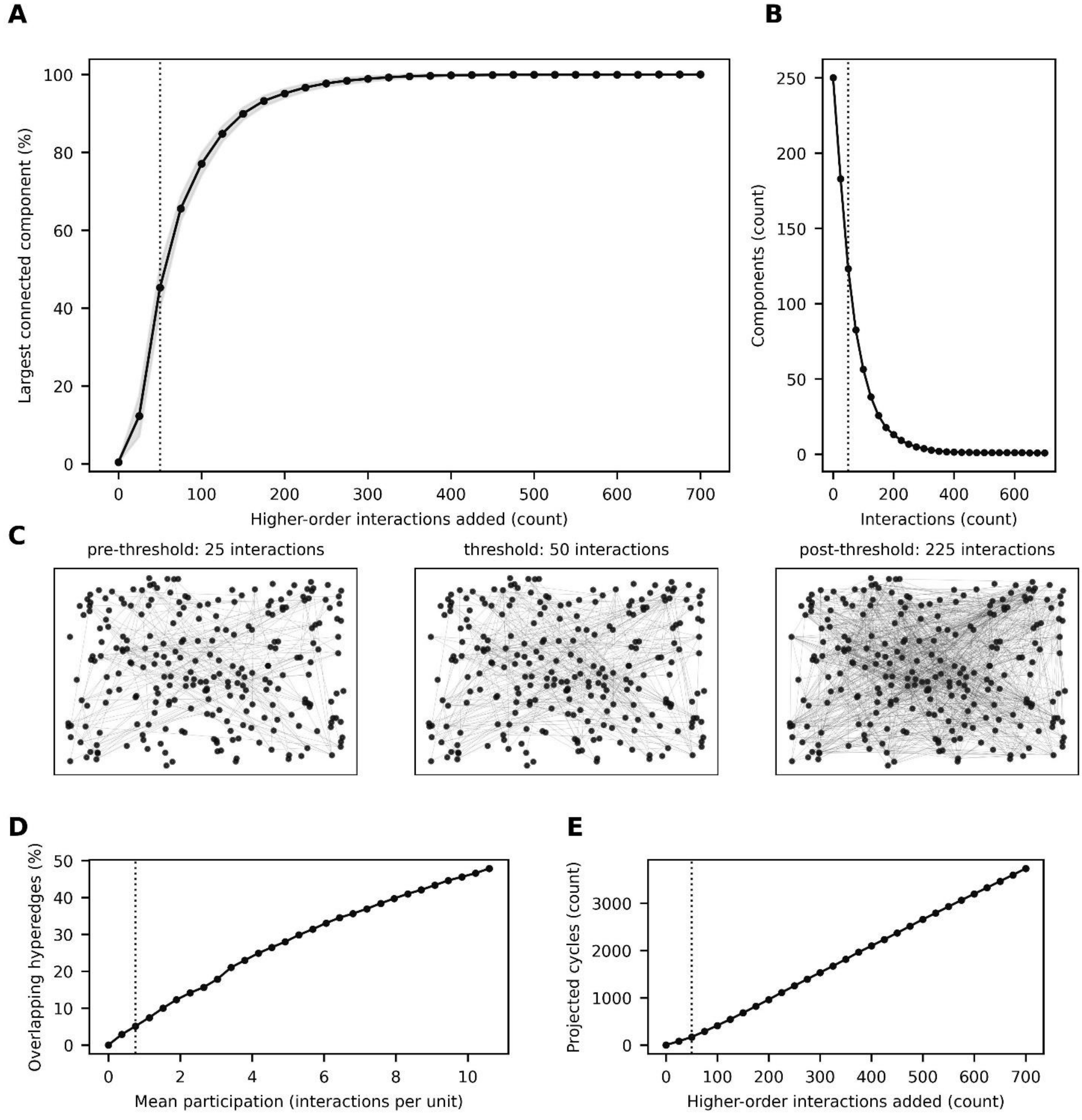

**Figure 1**. Hypergraph percolation and rapid emergence of integrated organization. The figure summarizes the behavior of a simulated higher-order interaction system composed of 250 elementary units connected through hyperedges containing between two and six units. The objective was to examine how increasing densities of higher-order interactions influence the formation of large-scale integrated structures.

Inset A. Percentage of units belonging to the largest connected component as a function of the cumulative number of higher-order interactions introduced into the system. The shaded envelope represents variability across 35 independent simulation replicates. A marked increase in component size occurs near the estimated transition region, indicating the rapid emergence of system-wide connectivity.

Inset B. Mean number of disconnected components in the system during the same process. The progressive reduction in component count reflects the merger of initially isolated assemblies into increasingly larger structures. The vertical dashed line denotes the estimated transition point, corresponding to approximately 50 higher-order interactions or roughly 0.75 interactions per unit.

Inset C. Representative projected network reconstructions before, during and after the transition region. Nodes correspond to elementary units and edges to pairwise projections derived from higher-order interactions. The left network illustrates a fragmented regime characterized by numerous disconnected clusters. The central network corresponds to the transition region, where multiple local assemblies become interconnected. The right network depicts the post-transition regime, in which a large integrated structure spans most units in the system.

Inset D. Percentage of overlapping hyperedges as a function of mean participation, measured as the average number of interactions associated with each unit. Increasing overlap reflects the formation of shared organizational structures linking previously independent assemblies.

Inset E. Number of projected cycles generated during network growth. Cycles provide a measure of structural redundancy and alternative pathways within the interaction network. Their rapid accumulation after the transition indicates increasing organizational reinforcement and connectivity.

Collectively, these panels illustrate how a relatively small increase in higher-order interaction density can generate an abrupt transition from fragmented local organization to extensive integrated structure through hypergraph percolation dynamics.

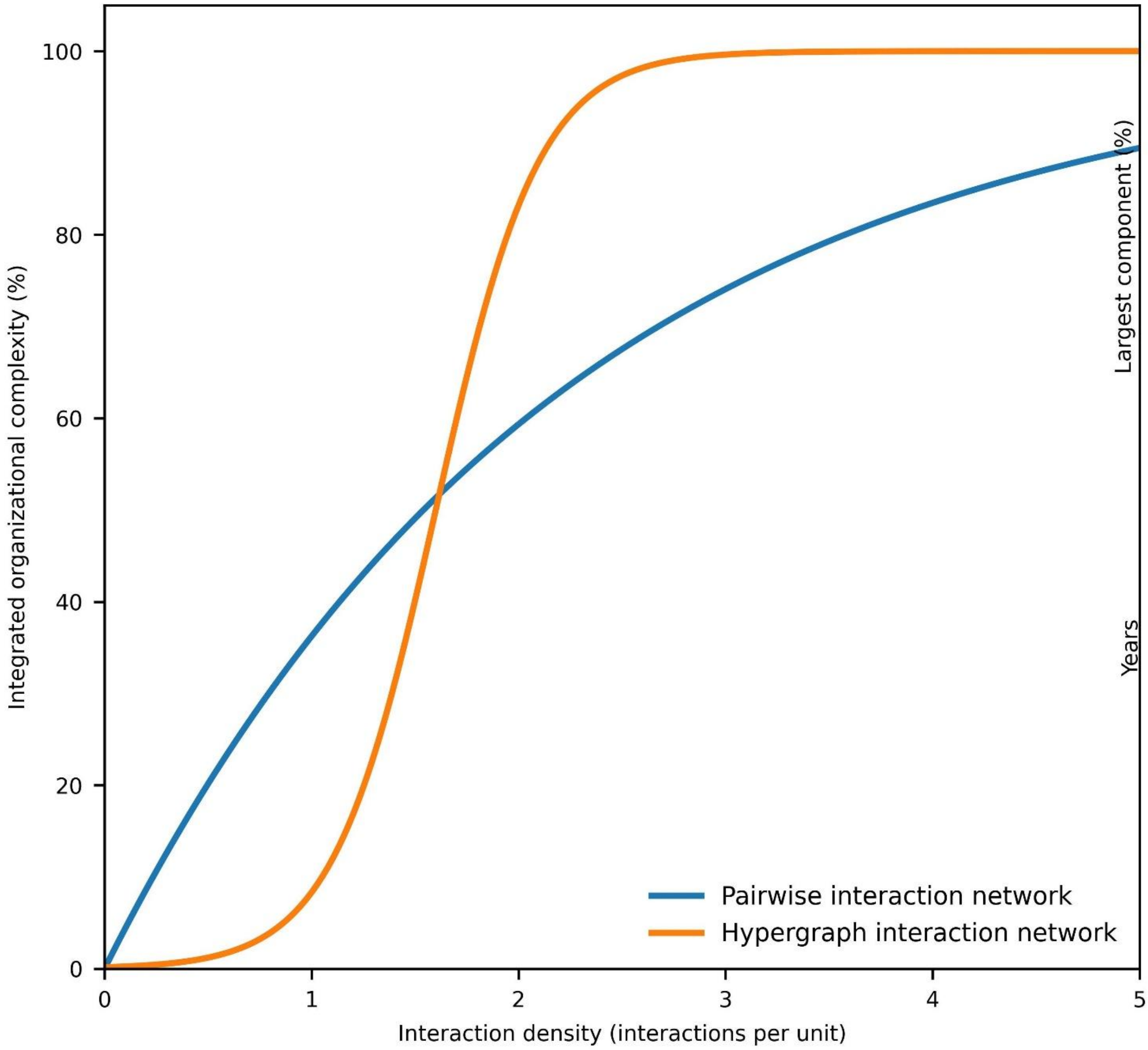

**Figure 2.** Hypergraph percolation accelerates the emergence of large-scale biological organization. The growth of integrated organizational complexity is compared in systems governed by pairwise interactions and higher-order hypergraph interactions. Complexity is quantified as the percentage of elementary units incorporated into the largest connected organizational assembly as interaction density increases. The pairwise model exhibits a gradual increase in connectivity, whereas the hypergraph model displays a sharp transition associated with the formation of extensive higher-order connectivity. Inset A enlarges the transition region and shows the growth of the largest connected component as a function of interaction density.

Therefore, higher-order interactions can generate abrupt increases in large-scale organization through hypergraph percolation dynamics.

CONCLUSIONS

We asked whether the emergence of complex biological organization can be described as a transition in higher-order interaction structure rather than exclusively as the accumulation of specific molecular innovations. We examined whether collective organization may arise when interactions involving multiple elements become dense enough to generate extensive connected assemblies. Our simulations compared systems built from progressively increasing higher-order interactions and quantified the resulting evolution of connectivity, overlap, participation, component structure and redundancy. We found that large-scale organization did not emerge through a smooth and approximately linear increase in connectivity. Instead, our simulations exhibited a localized transition region in which the largest connected component expanded rapidly while fragmentation declined. This transition was accompanied by increasing overlap among interactions and accumulation of alternative pathways within the projected interaction structure. These observations suggest that large changes in global organization can occur over relatively narrow intervals of interaction density. Our results support the interpretation that organizational transitions may complement molecular and biochemical explanations of complexity. Therefore, the emergence of integrated organization could depend not only on the presence of interacting elements, but also on the way interactions are distributed and interconnected throughout a system.
Research on biological complexity relies on different approaches. Autocatalytic approaches investigate the conditions under which reaction systems become self-sustaining through catalytic closure (Steel et al. 2019; Ravoni 2021; Hordijk et al. 2022; Xavier and Kauffman 2022; Loutchko 2023; Huson et al. 2024; Deng et al. 2026). Systems chemistry examines the emergence of collective properties from interacting molecular populations, while network-based descriptions represent biological systems as graphs composed of pairwise interactions. Percolation theory has been applied to reaction networks, ecological systems and connectivity transitions, whereas dynamical systems approaches focus on attractors, bifurcations and stability properties (Mello et al., 2021; Bianconi and Dorogovtsev, 2024; Cirigliano et al, 2024). Hypergraph representations differ from these techniques by treating interactions involving multiple elements as fundamental objects rather than reducing them to collections of pairwise links. This distinction becomes important whenever collective interactions cannot be decomposed naturally into independent binary relationships. Unlike graph-based connectivity measures, hypergraph connectivity retains information concerning the size and composition of collective interactions. Unlike catalytic closure models, our apporach is not restricted to particular chemical mechanisms. Unlike dynamical-system descriptions requiring explicit kinetic equations, our simulations focus on organizational structure generated by interaction topology, emphasizing how changes in interaction architecture could influence the emergence of extensive connectivity.

Our study has limitations. We used synthetic hypergraphs generated from prescribed probability distributions, not reproducing the detailed chemistry of prebiotic environments. Hyperedge sizes were restricted to interactions involving two to six elements, whereas real systems may contain broader and context-dependent interaction distributions. Connectivity was evaluated through graph projections for several observables, which simplify aspects of higher-order structure. The estimated transition region depends on the chosen node number, interaction-size distribution and interaction-accumulation process. Consequently, our reported thresholds should be interpreted as properties of the simulated systems rather than real constants. Additional sources of uncertainty arise from finite-size effects and from the possibility that alternative interaction-generation rules may produce different transition locations. Also, our simulations do not include energetic constraints, reaction kinetics, spatial heterogeneity, degradation processes or environmental fluctuations. Several questions are unresolved, e.g., the dependence of transition behavior on hyperedge-size distributions, the influence of temporally varying interactions, the relationship between hypergraph percolation and chemical reaction kinetics and the existence of universal scaling laws across different higher-order systems.

Our simulations suggest experimentally testable hypotheses. First, systems characterized by higher-order interactions should exhibit a critical interaction density at which the largest connected component increases more rapidly than expected from pairwise interaction models. This prediction can be tested in synthetic chemical networks, autocatalytic reaction systems or computational reaction ensembles by measuring the fraction of participating elements as interaction density increases.
Second, the percentage of overlapping interactions is predicted to increase sharply near the connectivity transition. Experimental measurements could quantify the frequency with which distinct reaction groups share common molecular participants and assess whether overlap follows a threshold-like trajectory.
Third, the number of alternative pathways, represented here by projected cycle counts, is expected to scale superlinearly after the transition region. Quantitative measurements of pathway redundancy could evaluate whether the number of independent routes between functional units increases faster than interaction density itself.
Fourth, transition locations should depend on hyperedge-size distributions. Systems dominated by larger collective interactions are predicted to require lower interaction densities to achieve extensive connectivity than systems dominated by pairwise interactions.
Fifth, finite-size scaling analyses can test whether transition sharpness changes systematically with system size. If the largest connected component is denoted by $S$, the node number by $N$ and interaction density by $\rho$, one may examine whether relations of the form $S \sim N^{\alpha}$ or $S \sim (\rho - \rho_c)^{\beta}$ emerge near the transition. Future studies may combine hypergraph percolation with explicit chemical kinetics, energetic constraints, stochastic environmental variation, spatial organization and empirical higher-order interaction datasets to evaluate the robustness of these predictions.
Our approach may also be relevant for investigations of complex interacting systems. Higher-order interaction descriptions can be used to characterize collective processes difficult to represent through pairwise networks alone. Potential areas of interest include biochemical reaction ensembles, microbial consortia, ecological interaction webs, neural populations and social systems containing group-level interactions. Hypergraph representations provide quantitative measures of overlap, participation, interaction multiplicity and large-scale connectivity that can be monitored as systems evolve. These measures may facilitate comparative analyses among systems that differ in composition but share common organizational features. Interaction-density estimates, overlap statistics and connectivity trajectories can also provide standardized descriptors for evaluating the structural evolution of higher-order systems across different observational scales.

In conclusion, we examined whether increasing densities of higher-order interactions could generate transitions from local organization to extensive connectivity. Our simulations showed coordinated changes in component size, fragmentation, overlap and redundancy during hypergraph growth. This suggests that interaction architecture and higher-order connectivity could provide a quantitative perspective for analyzing organizational transitions in systems composed of interacting elements, including the transition from prebiotic assemblies to early biological organization.

## DECLARATIONS

**Ethics approval and consent to participate.** Ethics approval statement is not applicable. This research does not contain any studies with human participants or animals performed by the Author.
**Consent for publication.** The Author transfers all copyright ownership, in the event the work is published. The undersigned author warrants that the article is original, does not infringe on any copyright or other proprietary right of any third part, is not under consideration by another journal and has not been previously published.
**Availability of data and materials.** All data and materials generated or analyzed during this study are included in the manuscript. The Author had full access to all the data in the study and took responsibility for the integrity of the data and the accuracy of the data analysis.
**Disclaimer**. The views expressed are those of the author and do not necessarily reflect those of the affiliated institutions.
**Competing interests.** The Author does not have any known or potential conflict of interest including any financial, personal or other relationships with other people or organizations within three years of beginning the submitted work that could inappropriately influence or be perceived to influence their work.
**Funding.** This research did not receive any specific grant from funding agencies in the public, commercial or not-for-profit sectors.
**Acknowledgements:** none.
**Authors' contributions.** The Author performed: study concept and design, acquisition of data, analysis and interpretation of data, drafting of the manuscript, critical revision of the manuscript for important intellectual content, statistical analysis, obtained funding, administrative, technical and material support, study supervision.
**Declaration of generative AI and AI-assisted technologies in the writing process.** During the preparation of this work, the author used ChatGPT 5.3 to assist with data analysis and manuscript drafting and to improve spelling, grammar and general editing. After using this tool, the author reviewed and edited the content as needed, taking full responsibility for the content of the publication.

## REFERENCES

1) Bianconi, G., and S. N. Dorogovtsev. 2024. "Theory of Percolation on Hypergraphs." Physical Review E 109 (1): 014306. https://doi.org/10.1103/PhysRevE.109.014306.
2) Brunk, N. E., and R. Twarock. 2021. "Percolation Theory Reveals Biophysical Properties of Virus-like Particles." ACS Nano 15 (8): 12988–12995. https://doi.org/10.1021/acsnano.1c01882.
3) Cirigliano, L., C. Castellano, and G. Bianconi. 2024. "General Theory for Extended-Range Percolation on Simple and Multiplex Networks." Physical Review E 110 (3): 034302. https://doi.org/10.1103/PhysRevE.110.034302.
4) Delabays, R., G. De Pasquale, F. Dörfler, and Y. Zhang. 2025. "Hypergraph Reconstruction from Dynamics." Nature Communications 16 (1): 2691. https://doi.org/10.1038/s41467-025-57664-2.
5) Deng, S., D. Lancet, and R. Yaniv. 2026. "Evolution from Composome to RNA Replicase." Life 16 (2): 219. https://doi.org/10.3390/life16020219.
6) Dharmaprani, D., E. V. Jenkins, K. Tiver, S. S. Shahrbabaki, C. Strong, D. Chapman, and A. N. Ganesan. 2023. "Percolation Theory as a Conceptual Framework to Explain Spontaneous Atrial Fibrillation Termination: A Pilot

Study.” Proceedings of the Annual International Conference of the IEEE Engineering in Medicine and Biology Society 2023: 1–4. https://doi.org/10.1109/EMBC40787.2023.10340363.
7) Feng, Y., J. Han, S. Ying, and Y. Gao. 2024. “Hypergraph Isomorphism Computation.” IEEE Transactions on Pattern Analysis and Machine Intelligence 46 (5): 3880–3896. https://doi.org/10.1109/TPAMI.2024.3353199.
8) Gao, Y., Y. Feng, S. Liu, X. Han, S. Du, Z. Wu, and H. Hu. 2026. “Hypergraph Foundation Model.” IEEE Transactions on Pattern Analysis and Machine Intelligence 48 (4): 4063–4080. https://doi.org/10.1109/TPAMI.2025.3647504.
9) Gao, Y., Z. Zhang, H. Lin, X. Zhao, S. Du, and C. Zou. 2022. “Hypergraph Learning: Methods and Practices.” IEEE Transactions on Pattern Analysis and Machine Intelligence 44 (5): 2548–2566. https://doi.org/10.1109/TPAMI.2020.3039374.
10) Hao, X., J. Li, M. Ma, J. Qin, D. Zhang, and F. Liu. 2024. “Hypergraph Convolutional Network for Longitudinal Data Analysis in Alzheimer’s Disease.” Computers in Biology and Medicine 168: 107765. https://doi.org/10.1016/j.compbiomed.2023.107765.
11) Hordijk, W., M. Steel, and S. Kauffman. 2022. “Autocatalytic Sets Arising in a Combinatorial Model of Chemical Evolution.” Life 12 (11): 1703. https://doi.org/10.3390/life12111703.
12) Huson, D., J. C. Xavier, and M. Steel. 2024. “Self-Generating Autocatalytic Networks: Structural Results, Algorithms and Their Relevance to Early Biochemistry.” Journal of the Royal Society Interface 21 (214): 20230732. https://doi.org/10.1098/rsif.2023.0732.
13) Ibarra-Junquera, V., D. Radillo-Ochoa, and C. A. Terrero-Escalante. 2022. “Evolutionary Timeline of a Modeled Cell.” Journal of Theoretical Biology 551–552: 111233. https://doi.org/10.1016/j.jtbi.2022.111233.
14) Kotlarz, P., J. C. Nino, and M. Febo. 2022. “Connectomic Analysis of Alzheimer’s Disease Using Percolation Theory.” Network Neuroscience 6 (1): 213–233. https://doi.org/10.1162/netn_a_00221.
15) Loutchko, D. 2023. “An Algebraic Characterization of Self-Generating Chemical Reaction Networks Using Semigroup Models.” Journal of Mathematical Biology 86 (5): 76. https://doi.org/10.1007/s00285-023-01899-4.
16) Lu, F., M. Ng, and A. Yip. 2026. “Hypergraph Neural Diffusion Networks.” Neural Networks 195: 108271. https://doi.org/10.1016/j.neunet.2025.108271.
17) Mallamace, F., G. Mensitieri, M. Salzano de Luna, P. Lanzafame, G. Papanikolaou, and D. Mallamace. 2022. “The Interplay between the Theories of Mode Coupling and of Percolation Transition in Attractive Colloidal Systems.” International Journal of Molecular Sciences 23 (10): 5316. https://doi.org/10.3390/ijms23105316.
18) Mathis, C., D. Patel, W. Weimer, and S. Forrest. 2024. “Self-Organization in Computation and Chemistry: Return to AlChemy.” Chaos 34 (9): 093142. https://doi.org/10.1063/5.0207358.
19) Mello, I. F., L. Squillante, G. O. Gomes, A. C. Seridonio, and M. de Souza. 2021. “Epidemics, the Ising-Model and Percolation Theory: A Comprehensive Review Focused on Covid-19.” Physica A 573: 125963. https://doi.org/10.1016/j.physa.2021.125963.
20) Miller, B. J. 2024. “A Percolation Theory Analysis of Continuous Functional Paths in Protein Sequence Space Affirms Previous Insights on the Optimization of Proteins for Adaptability.” PLoS ONE 19 (12): e0314929. https://doi.org/10.1371/journal.pone.0314929.
21) Notarmuzi, D., C. Castellano, A. Flammini, D. Mazzilli, and F. Radicchi. 2021. “Percolation Theory of Self-Exciting Temporal Processes.” Physical Review E 103 (2): L020302. https://doi.org/10.1103/PhysRevE.103.L020302.
22) Oliver, P., E. Zhang, and Y. Zhang. 2024. “Scalable Hypergraph Visualization.” IEEE Transactions on Visualization and Computer Graphics 30 (1): 595–605. https://doi.org/10.1109/TVCG.2023.3326599.
23) Ravoni, A. 2021. “Long-Term Behaviours of Autocatalytic Sets.” Journal of Theoretical Biology 529: 110860. https://doi.org/10.1016/j.jtbi.2021.110860.
24) Sanoria, M., R. Chelakkot, and A. Nandi. 2022. “Percolation Transition in Phase-Separating Active Fluid.” Physical Review E 106 (3): 034605. https://doi.org/10.1103/PhysRevE.106.034605.
25) Shin, D., J. Lee, J. R. Gong, and K. H. Cho. 2017. “Percolation Transition of Cooperative Mutational Effects in Colorectal Tumorigenesis.” Nature Communications 8 (1): 1270. https://doi.org/10.1038/s41467-017-01171-6.
26) Steel, M., W. Hordijk, and J. C. Xavier. 2019. “Autocatalytic Networks in Biology: Structural Theory and Algorithms.” Journal of the Royal Society Interface 16 (151): 20180808. https://doi.org/10.1098/rsif.2018.0808.
27) Unterberger, J., and P. Nghe. 2022. “Stoechiometric and Dynamical Autocatalysis for Diluted Chemical Reaction Networks.” Journal of Mathematical Biology 85 (3): 26. https://doi.org/10.1007/s00285-022-01798-0.
28) Xavier, J. C., and S. Kauffman. 2022. “Small-Molecule Autocatalytic Networks Are Universal Metabolic Fossils.” Philosophical Transactions of the Royal Society A: Mathematical, Physical and Engineering Sciences 380 (2227): 20210244. https://doi.org/10.1098/rsta.2021.0244.
29) Yang, Y., C. Zhang, H. Kim, and R. Chen. 2025. “Nucleation-Percolation Transition in Desiccation Cracking of Clay.” Physical Review E 111 (5): L053501. https://doi.org/10.1103/PhysRevE.111.L053501.